\documentclass[aps,prb,citeautoscript,twocolumn,superscriptaddress,floatfix]{revtex4}
\usepackage{amsfonts,amsmath,amsthm,amssymb,bm}
\usepackage{graphicx,graphics,color,float,multirow}
\usepackage{hyperref}

\begin{document}

\title{On the Statistical Mechanics of Mass Accommodation at Liquid-Vapor Interfaces}
\author{Kritanjan Polley}
\affiliation{Chemical Sciences Division, Lawrence Berkeley National Laboratory, Berkeley, CA 94720, USA}
\affiliation{Department of Chemistry, University of California, Berkeley, CA 94720, USA}
\author{Kevin R. Wilson}
\affiliation{Chemical Sciences Division, Lawrence Berkeley National Laboratory, Berkeley, CA 94720, USA}
\author{David T. \surname{Limmer}}
\email{dlimmer@berkeley.edu}
\affiliation{Chemical Sciences Division, Lawrence Berkeley National Laboratory, Berkeley, CA 94720, USA}
\affiliation{Department of Chemistry, University of California, Berkeley, CA 94720, USA}
\affiliation{Materials Sciences Division, Lawrence Berkeley National Laboratory, Berkeley, California 94720, USA}
\affiliation{Kavli Energy NanoScience Institute, Berkeley, California 94720, USA}

\begin{abstract}
We propose a framework for describing the dynamics associated with the adsorption of small molecules to liquid-vapor interfaces, using an intermediate resolution between traditional continuum theories that are bereft of molecular detail and molecular dynamics simulations that are replete with them. In particular, we develop an effective single particle equation of motion capable of describing the physical processes that determine thermal and mass accommodation probabilities. The effective equation is parameterized with quantities that vary through space away from the liquid-vapor interface. Of particular importance in describing the early time dynamics is the spatially dependent friction, for which we propose a numerical scheme to evaluate from molecular simulation. Taken together with potentials of mean force computable with importance sampling methods, we illustrate how to compute the mass accommodation coefficient and residence time distribution. Throughout, we highlight the case of ozone adsorption in aqueous solutions and its dependence on electrolyte composition.
\end{abstract}

\maketitle

\section{Introduction}\label{secIntro}
A statistical mechanical description of mass transport near liquid-vapor interfaces requires a detailed understanding of an environment that varies in physical and chemical properties over nanometer lengthscales.~\cite{rowlinson13,weeks77,garrett06,fumagalli18,limmer23} An extended interface breaks translational symmetry and imposes a spatial dependence to molecular properties,  altering the character of fluctuations in a way not easily anticipated.~\cite{kalos77,evans79,geissler13} Molecular dynamics simulations are capable of representing the emergent molecular properties at liquid-vapor interfaces, and have provided significant insight into the variation of  density and reactivity near them.~\cite{jungwirth2006specific,benjamin1996chemical,lbadaoui23,singh22,schile2019rate} Translating this insight into an understanding of experimental measurements of gas accommodation, however, is hampered by the wide range of timescales associated with the physical and chemical processes that determine it, precluding its study by straightforward simulation.~\cite{kolb10,abbatt12,von2020multiscale,cruzeiro22} We propose a reduced description of the molecular dynamics that affect gas adsorption by postulating and parameterizing an effective stochastic equation for the motion of a gas molecule in the vicinity of an interface. By developing a means of extracting the frictional forces that oppose small molecule motion from molecular dynamics simulations and using existing methods to extract the potential of mean force, we arrive at a description intermediate between phenomenological continuum equations and detailed molecular simulations. Using this description, we examine the dynamics of ozone at an air-water interface following its adsorption.  

The examination of gas uptake phenomena has been traditionally carried out with the so-called ``resistor model".~\cite{molina96,worsnop02,kolb02} The resistor model is a set of phenomenological continuum equations that decouples the factors influencing liquid-gas interactions and allows one to calculate overall uptake probabilities and the transformative gas-liquid collision rate with analytical expressions using an analogy to electrical resistance. These models are successful in limiting cases when molecular details can be ignored,~\cite{hanson97,ammann03,molina96,poschl07,vieceli2004uptake} as when bulk properties dictate the observable behavior. When applicable, the simplified forms of the resistor model provide compact expressions for the composition and thermodynamic state dependence of gas uptake. However, as they are phenomenological, the resistor model is not systematically improvable, and direct comparisons between experiments and simulations are complicated.~\cite{davidovits06} Explicit molecular dynamics simulations have been applied to the prediction of gas uptake.~\cite{vieceli05,julin13,neyertz06,galib21,hirshberg2018n} Thermal and mass accommodation probabilities can be computed with detailed molecular models, as can solubilities, though these calculations are expensive. Typically a resistor model framework is still used to transform these molecular calculations into a prediction of gas uptake. 

We aim to unify the molecular details of  explicit classical molecular dynamics simulation with the simplicity of the resistor model by developing a simple effective equation of motion for a gas molecule as it absorbs into a liquid.  To do this, we employ a single particle Fokker-Planck equation\cite{zwanzig2001nonequilibrium} for the center of mass position of a gas molecule relative to the liquid-vapor interface, generated from integrating out its intramolecular degrees of freedom and those of the surrounding solution.  Since the liquid-vapor interface breaks symmetry, to properly describe the gas molecule's dynamics we need to evaluate the position-dependent mean force and friction acting on the incoming particle. Standard free energy calculations~\cite{radmer97,klimovich15} can be used with molecular dynamics simulations to extract the potential of mean force. State-dependent friction has been considered in many instances, and some methods exist to extract its functional forms from simulation.~\cite{jayannavar95,dygas86,porra96,lanccon02,martin21,lau07} However, these methods typically work only in the condensed phase where the dynamics of the system are overdamped,~\cite{hummer05,woolf94} which is not applicable for a molecule in the gas phase. Here, we present an iterative method to obtain a spatially dependent friction near the liquid-vapor interface and use it within our reduced dynamical description.

Throughout, we use ozone at an air-water interface as an illustrative example of our framework. In particular, motivated by recent experimental observations we consider the effect of electrolyte composition in the liquid phase to the adsorption process.\cite{artiglia17,willis22,prophet24} The reactivity of ozone with iodide in aerosols at the marine boundary layer is of paramount importance in atmospheric chemistry.~\cite{vogt99,carpenter03,saiz12,carpenter13} A major fraction of overall ozone deposition from the marine boundary layer and emission of halogen gasses come from the reaction between the iodide and ozone.~\cite{ganzeveld09} Understanding how desorption, solvation, and adsorption are coupled together in this system are important questions for atmospheric chemistry and subjects of active research. While we used ozone at the air-water interface as a model system, our approach is general and applicable to any liquid-vapor interface and solute.

\section{Coarse-grained dynamics}\label{secCoarse}
\begin{figure}
    \centering
    \includegraphics[width=7.5cm]{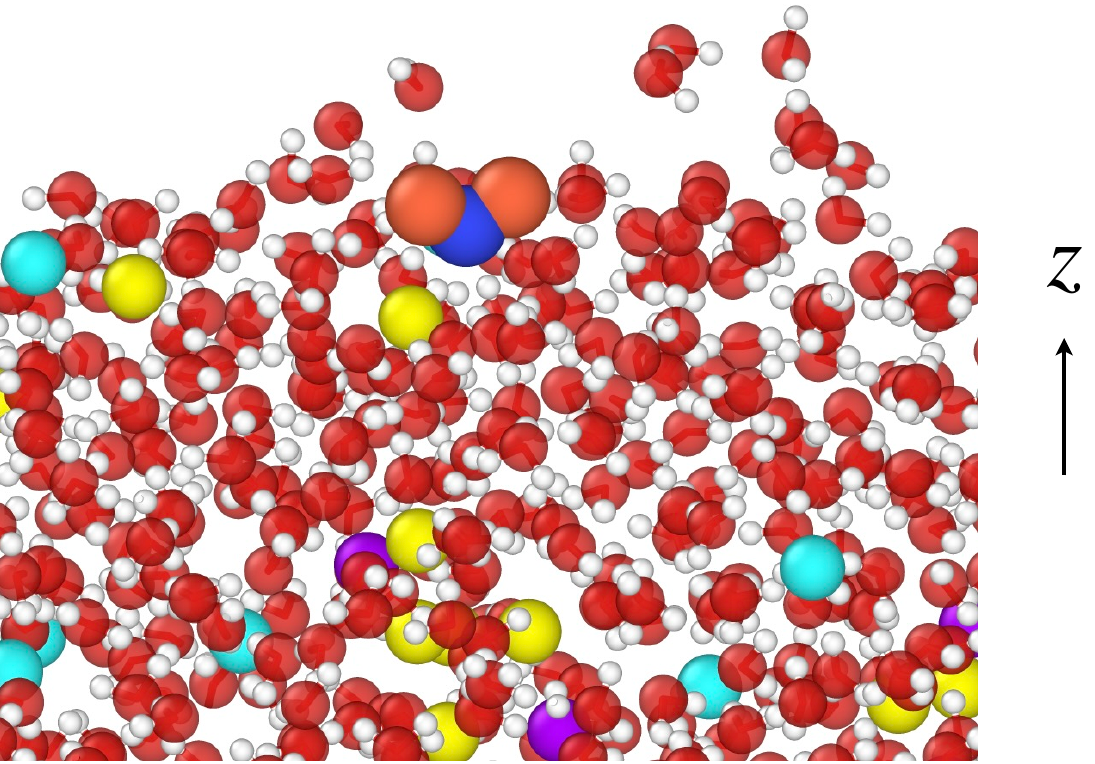}
    \caption{A snapshot of the MD simulation where an ozone molecule is near the air-water interface. The yellow, purple, and cyan particles are sodium, iodide, and bromide ions, respectively. The interface is perpendicular to the $z$-axis.}
    \label{figMDsnapshot}
\end{figure}

We aim to describe the dynamics of a small molecule initially in the gas phase, impinging upon a liquid-vapor interface, and subsequently either desorbing back into the vapor phase or solvating into the liquid. The level of description we choose to adopt is that intermediate between the full atomic resolution of molecular dynamics simulations and the macroscopic continuum equations of mass transport. In particular, we propose an effective equation of motion for the center of mass position and velocity of the small molecule. If the plane of the average liquid-vapor interface is oriented perpendicular to the $z$ axis, as illustrated in Fig.~\ref{figMDsnapshot}, then absent other sources of heterogeneity, there is translational symmetry in those directions, so it is sufficient to describe motion only in $z$. As we will consider only the motion of the small molecule center of mass, implicitly integrating out its internal degrees of freedom and those from constituents making up the liquid and vapor phases, our description is necessarily a statistical one.\cite{zwanzig2001nonequilibrium} Thus we aim to describe the probability, $p(z,v,t)$, of finding the small molecule at position $z$ with velocity $v$ at time $t$. Under additional assumptions that the solution is dilute, that the flux of gas particles onto the liquid interface is small, and that the resultant dynamics of the small molecule are Markovian, its distribution will satisfy a Fokker-Planck equation of the form,
\begin{align}\label{eqFPE}
    \frac{\partial p}{\partial t}= & -v\frac{\partial p }{\partial z}+ \frac{\partial}{\partial v} \left [ \frac{\partial_z F(z)}{m}+ \gamma(z) v \right ]p + \frac{k_{\mathrm{B}}T\gamma(z)}{m} \frac{\partial^2 p }{\partial v^2},
\end{align}
where $F(z)$ is the potential of mean force, and $\gamma(z)$ is the friction coefficient weighted by the molecule's mass $m$. The Markovian assumption requires we assume that the internal motions are fast relative to translational motion through the fluids and the dilute solution assumption that we ignore correlations between particles. We assume that the system is kept at constant temperature $T$, and $k_{\mathrm{B}}$ is Boltzmann's constant. The spatial symmetry breaking imposed by the extended liquid-vapor interface means that in general the friction and mean force are functions of the coordinate, $z$. However, away from the interface, into either the liquid or  vapor, both functions should approach constant values as the fluids regain local translational invariance. The distance over which the functions vary is reflective of correlations between molecules at the interface and those away from it. Apart from systems at critical points, it is expected that both $F(z)$ and $\gamma(z)$ decay to constants within a few molecular diameters of the interface. 

Equation ~\ref{eqFPE} encodes the framework we will use to connect the detailed description afforded by explicit molecular simulation, to a reduced description amenable to simplified numerical analysis and generalization. Traditional continuum model approaches can be recovered from this equation under additional assumptions of rapid velocity relaxation, rending the equation overdamped, and piecewise constant forms for  $F(z)$ and $\gamma(z)$. In principle both  $F(z)$ and $\gamma(z)$ can be inferred from experiment, but here they will be inferred directly from molecular dynamics simulations.

\section{Model of O$_3$ at an air-water interface}\label{secModel}

We adopted a polarizable classical force field for the molecular dynamics simulations of the air-water interface. A water slab with two air-water interfaces was used to represent an aqueous droplet containing 768 water molecules at 300 K in a box of size 24.8 $\times$ 24.8 $\times$ 111.8 \AA$^3$ where the larger dimension, denoted as the $z$-axis, is perpendicular to the interface. Periodic boundary conditions were applied in all directions. Water, ions and ozone were simulated with a polarizable force field in conjunction with SWM4-NDP~\cite{lamoureux06} as the water model. We employed rigid body dynamics for the water and ozone molecules following Ref.~\onlinecite{miller02} in order to use a time step of 1 fs. The non-bonded pair interactions were computed with a  Lennard-Jones potential
\begin{equation}
    U_{\text{LJ}} = \sum_{i,j} 4\epsilon_{ij}\left[ \bigg( \frac{\sigma_{ij}}{r_{ij}} \bigg)^{12} - \bigg( \frac{\sigma_{ij}}{r_{ij}} \bigg)^{6}  \right], \label{eqLJ}
\end{equation}
where $r_{ij}$ is the distance between atoms $i$ and $j$, while $\sigma_{ij}$ and $\epsilon_{ij}$ are Lennard-Jones parameters.  Table~\ref{tabLJparams} summarizes these parameters along with polarizability for the atoms. A Lorentz-Berthelot mixing rule was used where $\sigma_{ij}=(\sigma_{ii}+\sigma_{jj})/2$ and $\epsilon_{ij}=\sqrt{\epsilon_{ii}\epsilon_{jj}}$. A Drude oscillator model was used to replicate polarization in the simulation.~\cite{huang14,lemkul16} A spring constant, $k_\mathrm{D}$ of 1000 kcal/mol/\AA$^{2}$ was set for all Drude oscillators in the system and a charge $q_\mathrm{D}$ was choose so that the Drude particle produced the correct polarizability, $q_\mathrm{D} = -\sqrt{\alpha k_\mathrm{D}}$.~\cite{lamoureux03}

\begin{table}
    \renewcommand*{\tabcolsep}{6pt}
    \renewcommand*{\arraystretch}{1.2}
    \begin{center}
    \begin{tabular}{ c|c|c|c|c}
    \toprule
      Species & Atom & $\epsilon$(kcal/mol) &  $\sigma$(\AA)& $\alpha$(\AA$^3$)  \\
      \hline \hline
       \multirow{2}*{H$_2$O} & H                   & 0.0000 & 0.0000 & 0.0000 \\ 
                             & O                   & 0.2109 & 3.1839 & 0.9783 \\
        \multirow{2}*{O$_3$} & O$_{\text{center}}$ & 0.1560 & 3.2037 & 0.9500 \\
                             & O$_{\text{side}}$   & 0.1560 & 3.2037 & 0.9500 \\
        Na$^+$               & Na                  & 0.1000 & 2.2718 & 0.2400 \\
        Cl$^-$               & Cl                  & 0.1000 & 4.3387 & 3.6900 \\
        I$^-$                & I                   & 0.1000 & 5.1245 & 6.9200 \\
      \botrule
    \end{tabular}
    \caption{The force field parameters used in MD simulation. The fifth column ($\alpha$) represent the polarizability. The water force field parameters and geometry are taken from Ref.~\onlinecite{lamoureux06}, ozone force field from Ref.~\onlinecite{vieceli05} and the alkali halide force field from Ref.~\onlinecite{dang02}. For ozone molecule, $r_{\text{eq}}$(O$_{\text{center}}$-O$_{\text{side}}$)=1.28 \AA\, and $\theta_{\text{eq}}$=116.7$^{\circ}$.  The charge on the central O atom in ozone molecule is $+0.19|e|$ and the charge on the side O atoms are $-0.095|e|$, where $e$ is the charge of an electron.}
    \label{tabLJparams}
    \end{center}
\end{table}

An extended Lagrangian dynamics, with velocity-Verlet~\cite{frenkel01book} time integration scheme, was used in which a small mass and kinetic energy are attributed to the Drude particles. The amplitude of the Drude oscillator was controlled with a low temperature thermostat (1 K) acting in the local center-of-mass frame of each atom-Drude pair.~\cite{lamoureux03} Thole damping~\cite{noskov05} was used to modulate the electrostatic interaction between particles and induced dipoles. The atom-specific Thole factors introduced in this model provide fine-tuning of near-field electrostatics. A particle-particle-particle-mesh method~\cite{pollock96} was used for the long range Coulomb interactions with a target relative error of 10$^{-5}$. The force field was symmetrized with the procedure outlined by Dodin and Geissler.~\cite{dodin23} The Lennard-Jones interactions  were truncated and shifted at a distance of 12 \AA. The systems we have studied in this manuscript include O$_3$ in pure water, and two different electrolyte solutions one with 0.28 M NaI and one with 0.28 M NaI and 0.84 M NaCl. All systems were studied at 300 K.

\section{Potential of mean force}\label{secMD}
A quantity that enters into Eq. ~\ref{eqFPE} is the potential of mean force, $F(z)$. The potential of mean force, or equivalently the free energy change to reversibly move a particle from one position to another, is equal to the probability of finding that particle within an equilibrium ensemble. Thus, $F(z)$ is computable as
\begin{equation}
    F(z)=-k_{\mathrm{B}}T \ln \langle \delta(z-z_0)\rangle+F_0,\label{eqPMF}
\end{equation} 
where the angular bracket indicates a canonical average over the Dirac delta function for the center of mass coordinate of ozone, $z_0$. The added constant, $F_0$, is a reference free energy that we will set 0 in the gas phase.  The free energy for transferring one ozone molecule through the air-water interface for the different aqueous solutions can be computed within our molecular dynamics simulations using umbrella sampling.~\cite{frenkel2023understanding} In all simulations, a harmonic potential of the form $k_z(z_0-\bar{z})^2/2$ is added to the system with $k_z=4$ kcal/mol/\AA$^2$ being the spring constant of the bias centered at $\bar{z}$. The center of mass of ozone is computed relative to the center of mass of the water slab. A total of $71$ simulations with different $\bar{z}$ values ranging from $-35$ \AA \, to $35$ \AA\, equally spaced at an interval of $1$ \AA\, were used. The slab geometries were equilibrated for 1 ns followed by a production run for 15 ns. All statistical properties are evaluated from the final production run and averaged over all independent simulations. 

\begin{figure}[b]
    \centering
    \includegraphics[width=8.5cm]{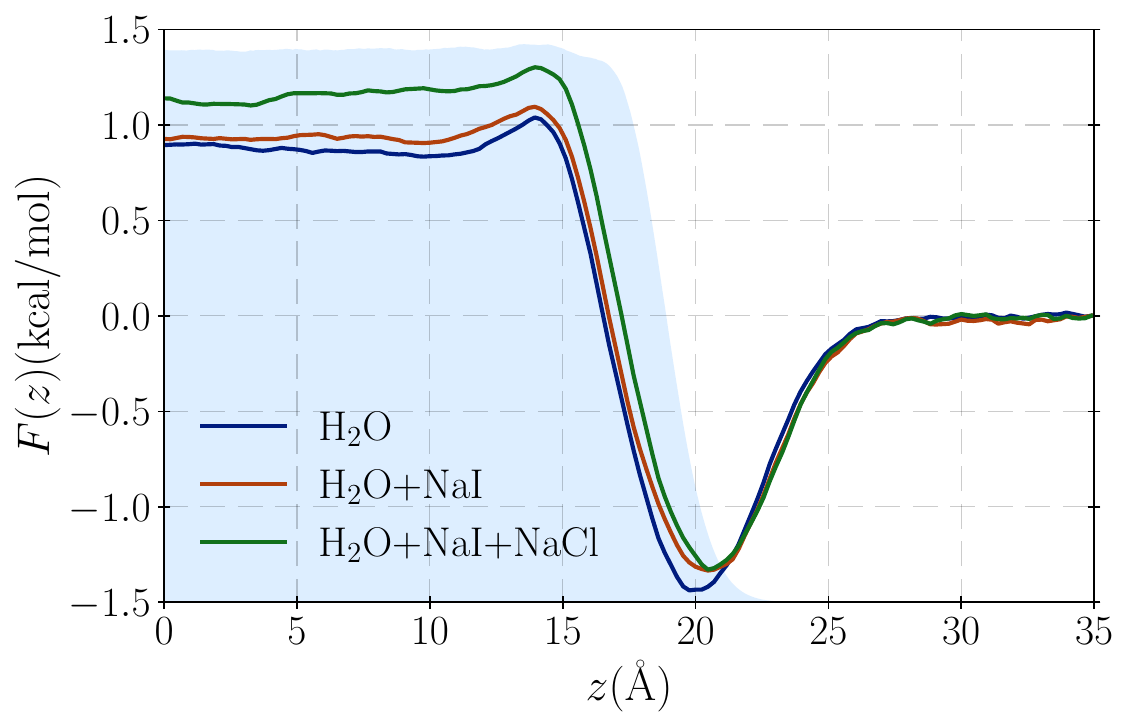}
    \caption{The free energy profiles for moving an ozone molecule through a variety electrolyte solutions. The shaded light blue region shows the location of the water density profile for reference.}
    \label{figPMF}
\end{figure}

The different simulations are used together with the weighted histogram analysis method~\cite{kumar92} to construct the free energy profiles. These are displayed in Fig.~\ref{figPMF}. The minima near the interface suggest the interface to be a thermodynamically favorable place for ozone molecules. The free energy plateaus far from the interface that sets a width for interfacial region of around $10$ \AA.  The solvation free energy ($\Delta F_{\mathrm{solv}}$), defined as the plateau value for ozone in liquid, for the pure water system of $0.86$ kcal/mol agrees nicely with experimental value of $0.84 \pm 0.07$ kcal/mol,\cite{vieceli05} which determines the dimensionless Henry's law constant ($ \exp[-\Delta F_{\mathrm{solv}}/k_{\mathrm{B}}T]= 0.236$). These results are similar to free energy calculation results from other molecular dynamics simulations.~\cite{vieceli05,vacha04,liyana11} The depth in the potential minima increases slightly with addition of salts as does the solvation energy, both of which are consistent with experimental observations.~\cite{benjamin91,kosak83,panich07} The destabilization at the interface in electrolyte solutions, compared to pure water system, is a consequence of iodide's strong preference for the interface. The propensity of iodide for the interface has been proposed to originate from an interplay between dielectric response and competing forces of solvation due to volume exclusion and dielectric polarization.~\cite{noah09,devlin22,jubb12,otten2012elucidating}


\section{Position Dependent Friction}\label{secAlgo}
The friction of ozone strongly depends on its position relative to the air-water interface. Asymptotically, its diffusion constant, given by the Einstein relation $D=k_\mathrm{B} T /m\gamma$, in bulk liquid is approximately 4 orders of magnitude smaller than in the gas phase,~\cite{wilson22} so $\gamma(z)$ must interpolate between these two limits over the molecular width of the liquid-vapor interface. While in homogeneous fluids the Einstein relation can be used to extract the friction, these are not possible to use to extract its spatial dependence.  Generalized Green-Kubo expressions have been developed to compute spatial-dependent mobilities and diffusivities in confined fluids,\cite{domingues23,mangaud2020sampling,helms2023intrinsic} that could work in principle to extract $\gamma(z)$, however the calculation of such correlation functions in the gas phase is difficult due to the rare collisions that dominate its decay. 

In the overdamped limit, many methods have been proposed to compute position dependent friction, often in biased simulations like those used for umbrella sampling, by fitting timeseries data to an effective Markov model, or Ornstein-Uhlenbeck process.~\cite{comer13,ljubetivc14,sicard21,bullerjahn20,marrink94,marrink96,gaalswyk16,lee16,breoni21,hummer05,woolf94} However, the overdamped assumption of these methods clearly breaks down in the vapor side of the interface. Liu et al. proposed a method to elucidate diffusion tensor for inhomogeneous systems such as gas-liquid interfaces with a known free energy surface.~\cite{liu04} The friction coefficient is computed by matching the survival probability from generalized Langevin dynamics using an exponential memory to molecular dynamics simulations by imposing various boundary conditions on the molecular system. 
This procedure becomes hard to converge in the low friction regime. A modification to this method with a Bayesian approach applied to Lennard-Jones type fluids was suggested by Colmenares et al., which allows one to bypass the numerical propagation of Langevin dynamics.~\cite{colmenares09}  We will adopt a similar approach, but one that works within a Markovian approximation and is quickly convergent even in the gas phase.

\subsection{Fitting $\gamma(z)$ from simulation}
\begin{figure}[b]
    \centering
    \includegraphics[width=8.5cm]{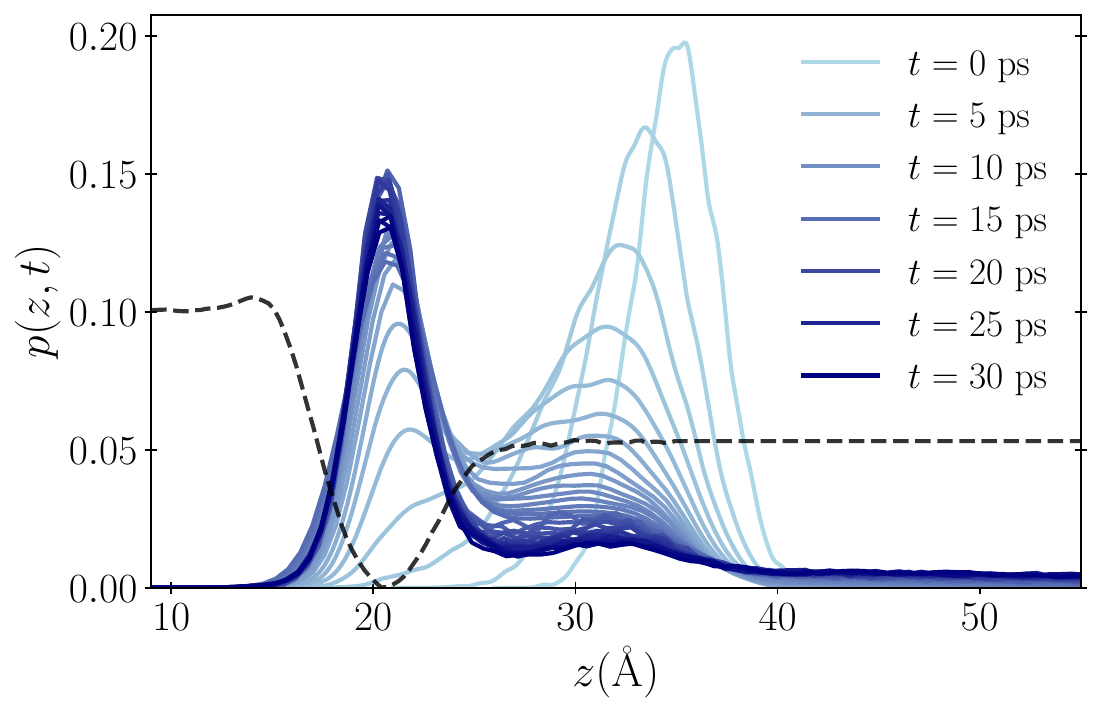}
    \caption{Evolution of position distribution of ozone obtained from molecular dynamics simulation for the system with NaI and NaCl.  The darker curves correspondent to later times. The dashed black curve (scaled and shifted) shows the free energy profile for this system.}
    \label{figMd_density_water}
\end{figure}

In order to extract the friction profile from molecular dynamics simulations, we apply a fitting procedure to match the results of a simulation with that predicted by Eq.~\ref{eqFPE}. In particular, we constructed an ensemble of initial conditions, $p(z,v,0)$, consisting of an ozone molecule in the gas phase with position $\bar{z}$ with a $z$-component center of mass velocity, $v_0$, pointed towards the slab of water,
\begin{equation}
p(z,v,0) = 2 \Theta (-v)\delta(z-\bar{z})  \left \langle  \delta (v-v_0) \delta (z-z_0) \right \rangle,
\end{equation}
where $\Theta (v)$ is the Heaviside step function, and all of the other degrees of freedom not constrained are averaged within a canonical ensemble. The delta function in $z$ is approximated by a narrow Gaussian. We then computed a time-dependent distribution, conditioned on $p(z,v,0)$,
\begin{equation}
p_\mathrm{MD}(z,v,t) = \left \langle  \delta [v-v_0(t)] \delta [z-z_0(t)] \right \rangle_{p(z,v,0)},
\end{equation}
where the subscript on the average implies the conditioning. We choose $\bar{z}=35$ \AA, and approximate the distribution by generating  approximately 15000 trajectories. We recorded the location and velocity of the center of mass of ozone along $z$ as a function of time as it approached the interface. The position probability distribution is obtained by integrating over velocities, 
\begin{equation}
p(z,t)=\int \mathrm{d}v \, p(z,v,t),
\end{equation}
and are shown for the system with water and NaI and NaCl in Fig.~\ref{figMd_density_water}.  At long times the position distribution 
 concentrates near the minima of the free energy profile. The shoulder at larger times originates from the decaying amplitude of the initial distribution and the scattering trajectories from the surface as the distribution moves towards the minima near the interface.

We then try to fit the data generated from the molecular dynamics simulations using the Fokker-Planck equation by varying the functional form of the friction. We assume a hyperbolic tangent shape for the friction function which has a form
    \begin{align}
         \gamma(z,\bm{\theta}) = &\Delta \gamma \Big(\tanh \big(w(s-z)\big)/2 + 1/2 \Big)^n  + \gamma_g, \label{eqGammaZ}
    \end{align}
    where $\Delta \gamma = \gamma_l - \gamma_g$ and $\gamma_g$ and $\gamma_l$ are friction of ozone in the gas phase and bulk liquid phase, respectively, while $\bm{\theta} \equiv \{w, s, n\}$ are fitting parameters. This functional form has the correct asymptotics, and could be relaxed to a more flexible ansatz though we found this unnecessary for the present systems. We use an initial condition for the Fokker-Planck equation, 
\begin{equation}
p(z,v,0) = \frac{\Theta (-v)}{\pi\sqrt{k_\mathrm{B} T \xi ^2 /m} }  e^{-m v^2/2 k_\mathrm{B} T}  e^{- (z-\bar{z})^2/2 \xi^2},
\end{equation}
consistent with that used in the molecular dynamics simulations, where $\xi$ accounts for the finite width of the initial $z$ distribution. We solve Eq.~\eqref{eqFPE} on a 2D grid, with a 6$^{\text{th}}$ order central finite difference for derivative operators. A fourth order Runge-Kutta method was used to propagate Eq.~\eqref{eqFPE}, to obtain the probability density for position and velocity at a given time, $p_{\mathrm{FP}}(z,v,t)$. We then define a loss function $\mathcal{L}(\bm{\theta})$,
\begin{equation}
    \mathcal{L}(\bm{\theta}) =  \int \bm{\mathrm{d}\Omega} \, \Big| p_{\mathrm{FP}}(z,v,t) - p_{\mathrm{MD}}(z,v,t) \Big|^2, \label{eqloss}
\end{equation}
where $\bm{\mathrm{d}\Omega} \equiv \mathrm{d}t\mathrm{d}z\mathrm{d}v$. To find optimized parameters for the friction in Eq.~\eqref{eqGammaZ} we used simulated annealing,~\cite{goffe94} to minimize the loss function in Eq.~\eqref{eqloss}.

\begin{figure}[b]
    \centering
    \includegraphics[width=8.5cm]{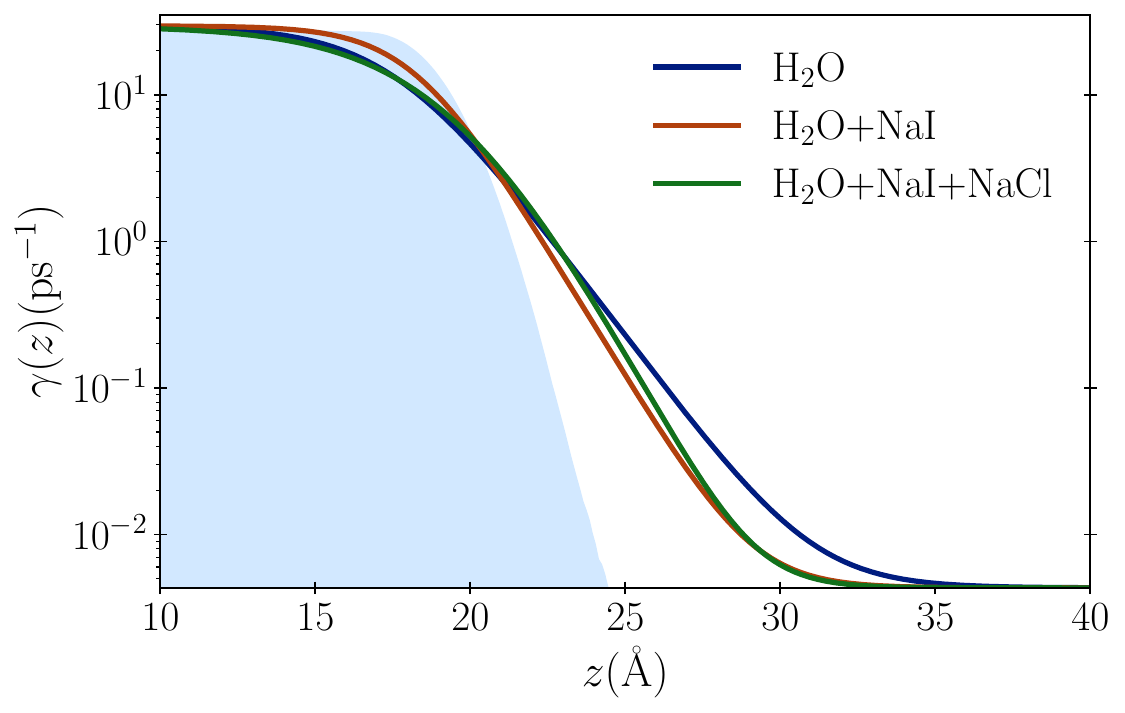}
    \caption{The optimized form of the friction functions for three systems. The $y$-axis is in log scale. The blue shaded region represents the water (scaled) density profile.}
    \label{figOptFriction}
\end{figure}
    
\begin{figure}[t]
    \centering
    \includegraphics[width=8.5cm]{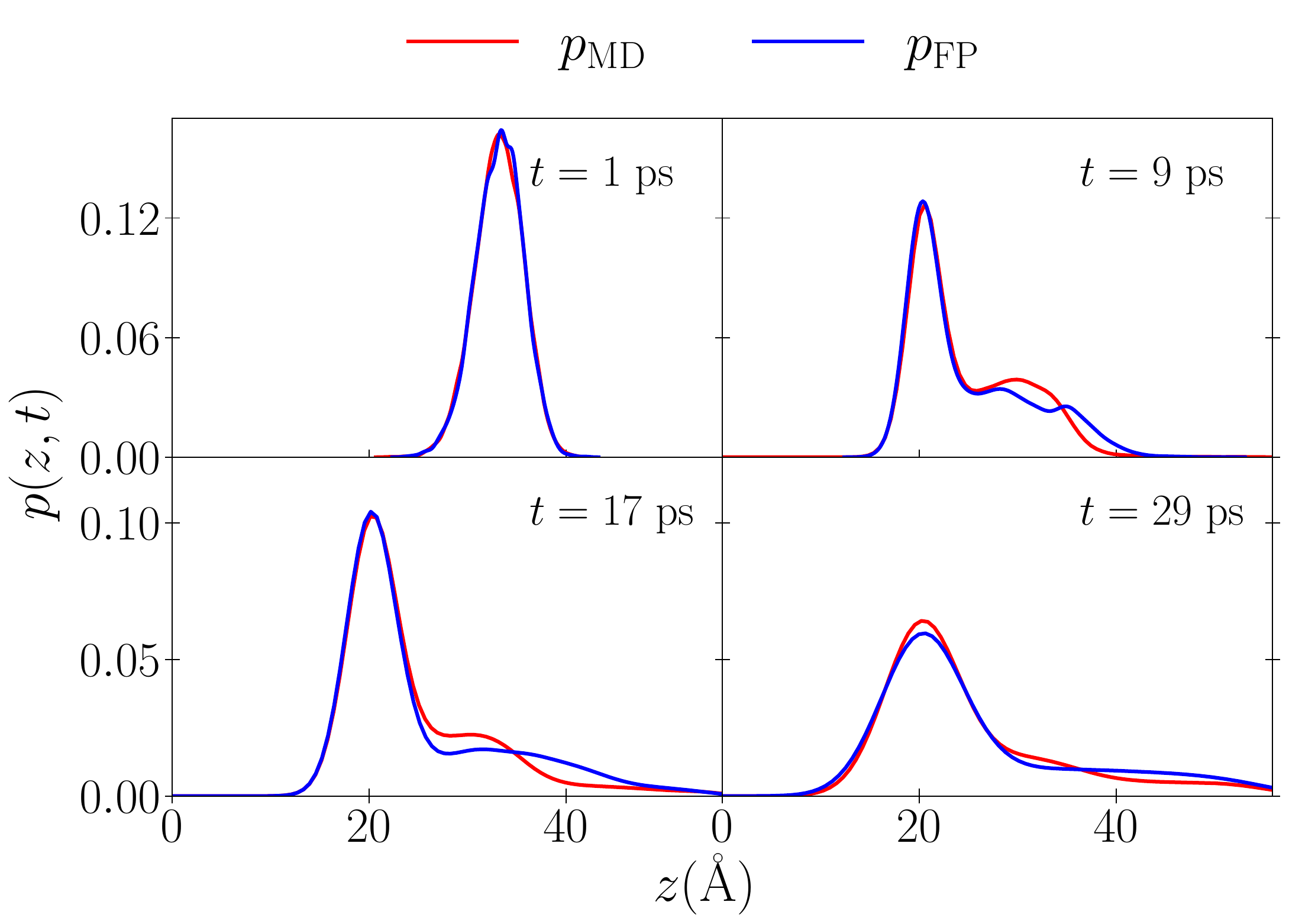}
    \caption{Comparison of position distribution obtained from shooting trajectories in molecular dynamics simulation and stochastic differential equation with optimized friction function. These results are for the system with ozone in pure water.}
    \label{figCompare_position_distribution}
\end{figure}

The optimized form of the friction functions are shown in Fig.~\ref{figOptFriction}. The friction of ozone in the gas phase and in the bulk liquid phase was computed using the Einstein relation, $\gamma_{l,g}=k_\mathrm{B} T /mD_{l,g}$ with $D_l=1.76\times 10^{-5} \, \mathrm{cm}^2/\mathrm{s}$ and $D_g=0.12 \, \mathrm{cm}^2/\mathrm{s}$.~\cite{wilson22} The bulk diffusion constant of ozone in solutions of sodium halides (within the concentration range used here) changes only slightly,~\cite{moreno19} and we have used the same bulk phase friction for all the systems studied here. The tails of the friction that rise before the ozone molecule reaches the interface manifests long range Coulomb interactions. The rise of the friction function for electrolyte solutions begins closer to the interface. This is a consequence of diminished capillary activity near the interfaces due to the presence of electrolytes. The initial forms of these friction functions were chosen to imitate the water density profile, the light blue shaded region in Fig.~\ref{figOptFriction}, though the optimize form deviates from it significantly. With the converged $\gamma(z)$, we are able to accurately reproduce $p_{\mathrm{MD}}(z,v,t)$, as illustrated in  Fig.~\ref{figCompare_position_distribution} for ozone in pure water. 
    
\section{Applications}\label{secApp}
With an ability to extract both $F(z)$ and $\gamma(z)$, we can fully parameterize the reduced dynamical description of ozone impinging upon an aqueous solution as encoded in our Fokker-Planck equation, Eq. ~\ref{eqFPE}. Using this expression, we have studied the dependence of ozone mass accommodation and interfacial residence time distribution on the electrolyte composition. 

\subsection{Mass Accommodation Coefficient}\label{subsecMAC}
The mass accommodation coefficient, $\Phi$, is defined as the fraction of collisions of the gas phase species with the interface that results in the transport of the gas phase particle into the condensed phase.~\cite{davidovits06} In the limit that solvation and desorption are both activated process, the mass accommodation coefficient can be computed as
\begin{equation}
    \Phi = \frac{k_{\text{solv}}}{k_{\text{des}}+ k_{\text{solv}}} \label{eqPhi},
\end{equation}
where $k_{\text{des}}$ and $k_{\text{solv}}$ are the rate of desorption and solvation, respectively, from the interface for the gas phase particle. This expression is valid in the limit of unit sticking coefficient and rapid thermalization of velocity relative to the absorption and desorption processes, both of which are valid for ozone on aqueous solutions.

Evaluating $\Phi$ from molecular simulations directly is complicated by the large activation barriers for both processes. Under the assumption that Eq. ~\ref{eqFPE} accurately describes the interfacial dynamics of ozone, we can use it to compute $\Phi$ at a minimal computational cost. While we could proceed by solving the Fokker-Planck equation directly within a basis, an alternative procedure is to propagate an equivalent Langevin equation,
\begin{equation}
    m\dot{v} = -m\gamma(z)v -\partial_z F(z) + \sqrt{2m\gamma(z)k_{\mathrm{B}}T}\eta(t) \label{eqLang},
\end{equation}
where the dot indicates a time derivative, $\eta(t)$ is the $\delta$-correlated Gaussian noise with $\langle \eta(t) \rangle = 0$, and $\langle \eta(t') \eta(t) \rangle = \delta(t-t')$. The diffusive dynamics for a particle with a spatially dependent friction requires extra care, as it implies a spatial dependence of the noise strength. These phenomena, for both underdamped and overdamped limit, have been studied extensively.~\cite{lau07,breoni21,baule08,farago14a,farago14b,regev16} We employ the methods detailed in Ref.~\onlinecite{farago14a} for our calculation that uses two different averages for dealing with coordinate dependent friction in Eq.~\eqref{eqLang} for the frictional force term and the noise term. They have shown that their approach is related closely to the Stratonovich interpretation~\cite{mannella12} and it produces the correct drift originating from the dissipation term and not from the noise term in the Langevin equation.

\begin{figure}
    \centering
    \includegraphics[width=8.5cm]{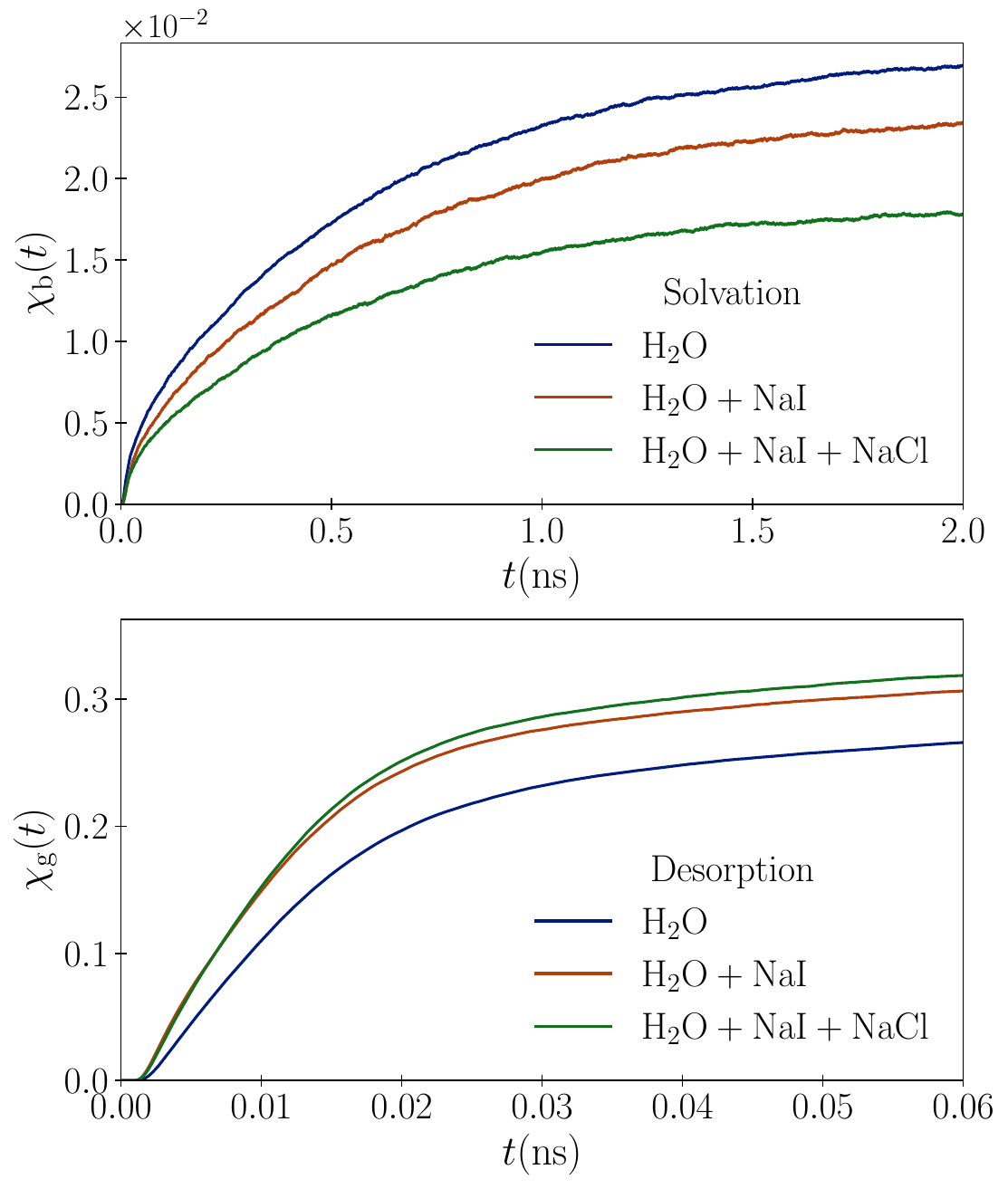}
    \caption{The side-side correlation functions, Eq.~\eqref{eqChi}, are plotted above. The solvation $\chi_\mathrm{b}(t)$, top and desorption $\chi_\mathrm{g}(t)$, bottom.}
    \label{figCorrRates}
\end{figure}

To determine the desorption and solvation rates, we compute the fraction of trajectories entering the gas phase or bulk phase as a function of time starting from the interfacial region. Indicator functions $h_\nu[z(t)]$,
\begin{align}
h_\nu[z(t)]  &= \begin{cases}
1 & \text{if } z(t)\in \nu=\{\mathrm{b,int,g}\} \\
0 & \text{otherwise} \\
\end{cases},
\end{align}
are used to determine the state an ozone molecule is in, being either bulk liquid phase ($\mathrm{b}$), interface ($\mathrm{int}$), or gas phase ($\mathrm{g}$). From these indicator functions, we can define side-side correlation functions,  $\chi_\nu(t)$, as
\begin{align}
\chi_\nu(t) & = \frac{\langle h_\nu[z(t)]h_{\mathrm{int}}[z(0)] \rangle}{\langle h_{\mathrm{int}}[z(0)] \rangle}, \label{eqChi}
\end{align}
which determines the fraction of trajectories that have entered or left a given region at a given time, given they started at the interface. The boundaries of the bulk solution and gas phase, are set where the free energy profile in Fig.~\ref{figPMF} becomes flat. We take the bulk as $z<14.5$, the interface as $14.5<z<27$ and the vapor as $z>27$. Rates for solvation or desorption are given by the time derivative of $\chi_\mathrm{b}$ or $\chi_\mathrm{g}$, respectively. Alternatively the rates can be determined by fitting the correlation functions with an exponential function of the form $\chi_\nu(t)=a_1-a_2e^{-kt}$ where $k$ is the rate for the process. Both of these procedures produce similar results. 

The correlation functions are plotted in Fig.~\ref{figCorrRates}. The initial condition for propagating Langevin trajectories was set at the minima of the free energy profile for position and a Boltzmann distribution in velocity. The inferred desorption rates match closely with expectations from detailed balance and kinetic theory. The ratio of solvation and desorption rates are obtained by propagating trajectories from the interface and placing absorbing boundary conditions in the liquid phase and in the vapor phase and counting the fraction of trajectories passing each absorbing boundary. From that ratio, we obtain consistent solvation rates. 

\begin{table}[htb!]
    \centering
    \begin{tabular}{c|c|c|c} 
    \toprule
    System &  $k_{\text{des}} \, (\text{s}^{-1})$ & $k_{\text{solv}} \, (\text{s}^{-1})$ & $\Phi(\times 10^2)$  \\
    \hline \hline
        O$_3$ (H$_2$O)           &  $1.40 \times 10^{10}$ & $2.66 \times 10^{8}$ & $1.18$  \\ 
        O$_3$ (H$_2$O+NaI)       &  $1.88 \times 10^{10}$ & $1.19 \times 10^{8}$ & $0.63$ \\
        O$_3$ (H$_2$O+NaI+NaCl)  &  $1.93 \times 10^{10}$ & $1.90 \times 10^{8}$ & $0.97$ \\
    \botrule
    \end{tabular}
    \caption{Desorption and solvation rates and the mass accommodation coefficient for ozone in different solutions}
    \label{tabRates}
\end{table}
The mass accommodation coefficients are computed from Eq.~\eqref{eqPhi}. The solvation and desorption rates and the mass accommodation coefficients are reported in Table~\ref{tabRates}. With 1500 explicit MD trajectories for desorption with ozone in pure water system, the desorption rate is obtained as $1.2 \times 10^{10}\, \mathrm{s}^{-1}$ which is close to the value obtained with the Langevin equation, confirming its applicability. The solvation rates are much smaller than desorption rates as the barrier for solvation is higher from the interface. The desorption rates for the three systems studied here are similar, but the solvation rates are significantly slower for the electrolyte solutions. This stems from a slightly higher barrier for solvation for the solution with I$^-$ relative to pure water, and a corresponding higher friction for the NaCl solution, implied slightly different mechanisms for solvation in both cases. The solvation and desorption rates observed here for ozone with NaI system were recently used in a kinetic model to successfully predict observed multiphase kinetics of iodide and ozone in microdroplets.~\cite{prophet24} 

\subsection{Residence Time Distribution}\label{subsecRTD}
The survival probability, $S(t)$, for a gas molecule at the interface dictates its ability to undergo interfacial specific processes like interfacial reactions. It can be obtained by integrating the probability density from Eq.~\eqref{eqFPE} over a restricted domain that measures only the remaining fraction of probability, $\widetilde{p}(z,v,t)$ in a given region. The probability, $\widetilde{p}(z,v,t)$,  differs from $p(z,v,t)$ due to absorbing boundary conditions outside the restricted domain that causes $\widetilde{p}(z,v,t)$ to loose probability. The survival probability is normalized such that at $t=0$, the value of the integral in Eq.~\eqref{eqSt} is 1. Specifically, the survival probability is
\begin{equation}
    S(t) = \int \mathrm{d}z \mathrm{d}v \,\, \widetilde{p}(z,v,t). \label{eqSt}
\end{equation}
and is computed using the same initial condition as in the calculation of $\Phi$, with absorbing boundary conditions in the bulk liquid and vapor phases.
The residence time distribution~\cite{berezhkovskii98,pury02} is the rate of loss of the survival probability, which is the negative time derivative of the above equation, $R(t)=-\partial S(t)/\partial t$. The residence time gives us insight into the inverse rate constant for diffusion controlled reactions where reactants are restricted in a finite domain. 

We seek the solution of Eq.~\eqref{eqFPE} with a complete basis set satisfying appropriate boundary conditions,
\begin{equation}
    \widetilde{p}(z,v,t)=\sum_{j,n} c_{j,n}(t) \zeta_j(z) \mu_n(v).\label{eqPzvt}
\end{equation}
We used sine functions as the spatial basis ($\zeta_j(z)$) and the Gaussian functions ($\mu_n(v)$) as the velocity basis function. The coefficients of the above equation, $c_{j,n}(t)$, are propagated according to Eq.~\eqref{eqFPE} using orthogonality of the sine functions and analytical Gaussian integrals.  
\begin{figure}
    \centering
    \includegraphics[width=8.5cm]{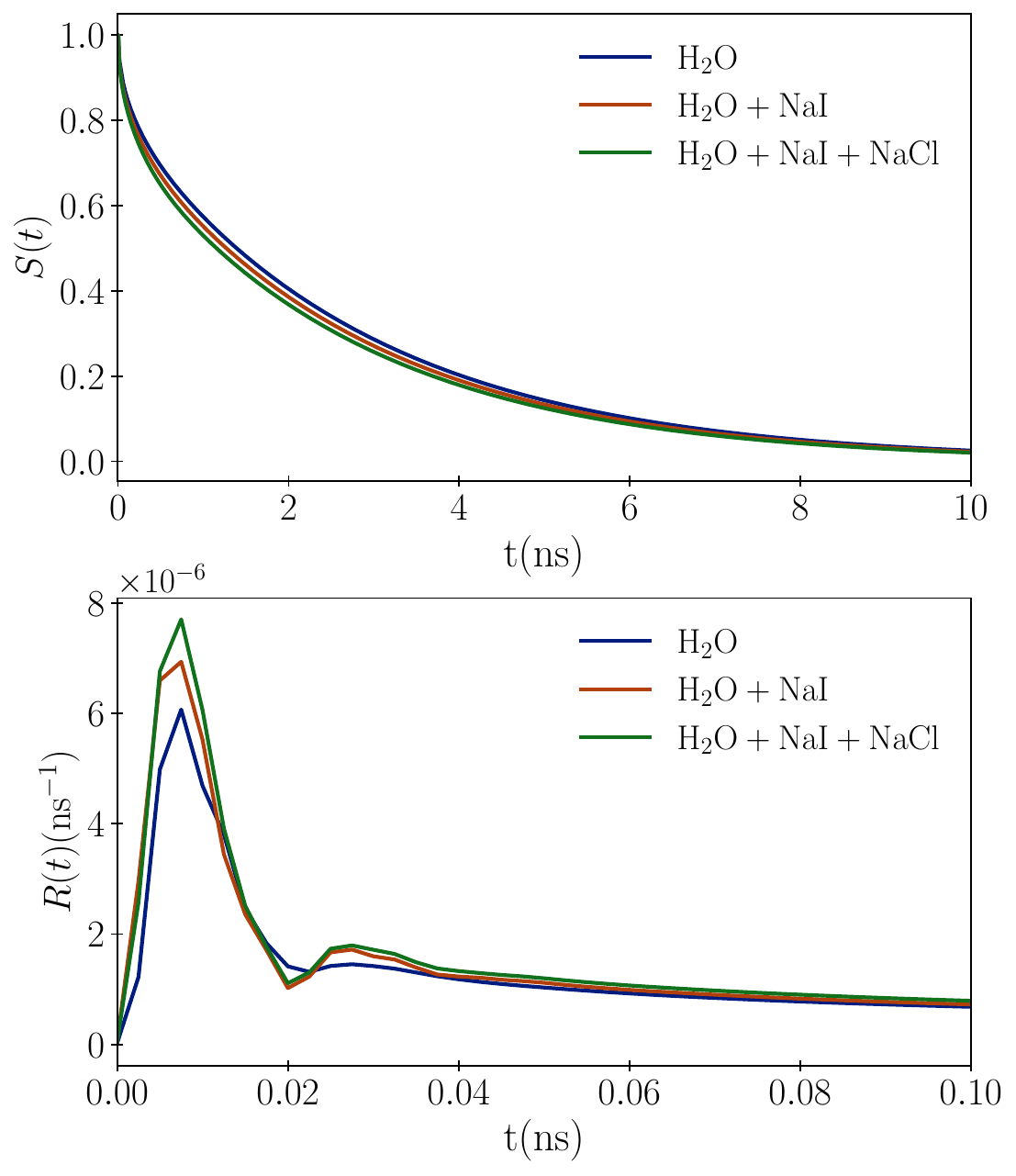}
    \caption{The top row shows the survival probability for ozone at the air-water interface. The bottom row depicts the residence time distribution obtained by taking the negative derivative of the plot above.}
    \label{figSt}
\end{figure}
The survival probability and residence time distribution profile are presented above in Fig.~\ref{figSt}. The early rise comes due to the fact that the initial condition in position is not an equilibrium distribution. The residence time distribution, $R(t)$, shows two distinct decay times, the first steeper decay corresponds to the faster desorption process, \textit{c.f.} Table~\ref{tabRates}, and the second slower one tallies with the slower solvation process. The broad distribution of timescales results from the two order of magnitude difference in desorption and solvation rates.

\section{Conclusion}\label{secconclusion}

In this article, we presented a way to integrate molecular details in stochastic differential equations to draw a continuous picture of the adsorption of solutes through liquid-vapor interfaces. The molecular details of the interface are incorporated into a stochastic equation of motion through a spatially dependent friction by optimizing it to reproduce molecular dynamics results and through free energy surfaces. Although we presented this framework for ozone near liquid-vapor interface, this framework can be generalized for any liquid vapor interfacial system. The form of friction function was chosen to be a hyperbolic tangent function to initially match the shape of the water density profile. Any other physically motivated form, and more complex function, can also be used with this algorithm. The free energy profile and friction function for a system carry the molecular signatures and, when included in a statistical picture, give us a better understanding of the system without getting into the every complex details of the fluctuations at the interfaces.~\cite{bedeaux85} This, in turn, provides a better handle for processes near liquid vapor interfaces like in aerosols and in marine-boundary layers~\cite{donaldson10,osthoff08,vogt99} where the high surface-to-bulk ratio of atmospheric aerosols makes their properties strongly dependent on the interfaces. We have applied this information to compute the mass accommodation coefficient and residence time distribution for different electrolyte solutions. The calculations of mass accommodation coefficients hand us direct quantitative knowledge of solvation and desorption rates. The theoretical framework reported here can be used as a starting point for a deeper understanding of more complex interfacial systems. Use of a reactive force field along with this methodology would further enable the calculation of reactive uptake, and finer insight into the kinetics and thermodynamics  of liquid-vapor interfaces.

{\bf Acknowledgments}
This work was supported by the Condensed Phase and Interfacial Molecular Science Program (CPIMS) of the U.S. Department of Energy under Contract No. DE-AC02-05CH11231. We sincerely thank Phill Geissler who supervised the earlier part of this project. We also thank Alex Prophet, Amr Dodin, and Liron Cohen for helpful discussions.

\bibliography{reference}

\end{document}